\documentclass[lettersize,journal]{IEEEtran}
\usepackage{cite}
\usepackage{amsmath,amssymb,amsfonts}
\usepackage{algorithmic}
\usepackage{array}
\usepackage[caption=false,font=normalsize,labelfont=sf,textfont=sf]{subfig}
\usepackage{textcomp}
\usepackage{stfloats}
\usepackage[hyphens]{url}
\usepackage{verbatim}
\usepackage{graphicx}
\usepackage{xcolor}

\usepackage{algorithm}
\usepackage{textcomp}
\usepackage{makecell}
\usepackage{multicol}
\usepackage{multirow}

\usepackage{color,soul}

\def\BibTeX{{\rm B\kern-.05em{\sc i\kern-.025em b}\kern-.08em
    T\kern-.1667em\lower.7ex\hbox{E}\kern-.125emX}}
\usepackage{balance}
\begin{document}
\title{MCAIMem: a \underline{M}ixed SRAM and eDRAM \underline{C}ell for \\Area and Energy-efficient on-chip \underline{AI} \underline{Mem}ory}
\author{Duy-Thanh Nguyen, Abhiroop Bhattacharjee, Abhishek Moitra, and Priyadarshini Panda \\
\{duy-thanh.nguyen, abhiroop.bhattacharjee, abhishek.moitra, priya.panda\}@yale.edu \\

Department of Electrical Engineering, Yale University, USA
\thanks{Manuscript created October, 2023; This work was developed by the Yale University. This work is distributed under the \LaTeX \ Project Public License (LPPL) ( http://www.latex-project.org/ ) version 1.3. A copy of the LPPL, version 1.3, is included in the base \LaTeX \ documentation of all distributions of \LaTeX \ released 2003/12/01 or later. The opinions expressed here are entirely that of the author. No warranty is expressed or implied. User assumes all risk.}}

\markboth{IEEE TRANSACTIONS ON VERY LARGE SCALE INTEGRATION (VLSI) SYSTEMS, VOL. 18, NO. 9, OCTOBER 2020}%
{How to Use the IEEEtran \LaTeX \ Templates}

\maketitle

\begin{abstract}
AI chips commonly employ SRAM memory as buffers for their reliability and speed, which contribute to high performance. However, SRAM is expensive and demands significant area and energy consumption. Previous studies have explored replacing SRAM with emerging technologies like non-volatile memory, which offers fast-read memory access and a small cell area. Despite these advantages, non-volatile memory's slow write memory access and high write energy consumption prevent it from surpassing SRAM performance in AI applications with extensive memory access requirements. Some research has also investigated eDRAM as an area-efficient on-chip memory with similar access times as SRAM. Still, refresh power remains a concern, leaving the trade-off between performance, area, and power consumption unresolved. To address this issue, our paper presents a novel mixed CMOS cell memory design that balances performance, area, and energy efficiency for AI memory by combining SRAM and eDRAM cells. We consider the proportion ratio of one SRAM and seven eDRAM cells in the memory to achieve area reduction using mixed CMOS cell memory. Additionally, we capitalize on the characteristics of DNN data representation and integrate asymmetric eDRAM cells to lower energy consumption. To validate our proposed MCAIMem solution, we conduct extensive simulations and benchmarking against traditional SRAM. Our results demonstrate that MCAIMem significantly outperforms these alternatives in terms of area and energy efficiency. Specifically, our MCAIMem can reduce the area by 48\% and energy consumption by 3.4$\times$ compared to SRAM designs, without incurring any accuracy loss.
\end{abstract}


\section{Introduction}
\label{sec:introduction}

\begin{figure}[!t]
\centering
\includegraphics[width=0.9\linewidth]{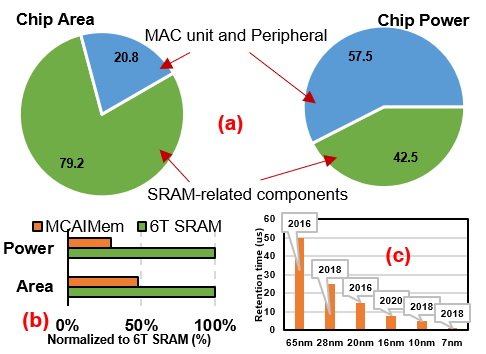}
\caption{(a) The breakdown of SRAM area and power in Eyeriss Chip~\cite{chen2016eyeriss} (Please note that, Eyeriss' FIFO is implemented by SRAM), (b) Our proposed MCAIMem can reduce 48\% area size and 3.4$\times$ power consumption compared to traditional 6T SRAM, and (c) eDRAM gain cell maintains its functionality even as the retention time diminishes at scaled technology nodes.}
\label{fig:srambreakdown}
\vspace{-2em}
\end{figure}

Deep Neural Network (DNN) accelerators have become crucial components in various machine learning systems. DNNs store a large number of parameters to achieve high accuracy, resulting in high memory requirements. DNNs have proven their effectiveness in a wide range of applications, including image recognition~\cite{krizhevsky2012imagenet}, object detection~\cite{huang2018yolo}, language translation~\cite{floridi2020gpt}, and autonomous driving~\cite{yurtsever2020survey}. State-of-the-art DNNs~\cite{floridi2020gpt} require billions of operations and a huge memory to store activations and weights, as evidenced by the 240$\times$ increase in transformer size over two years~\cite{gholami2020ai_and_memory_wall}. Dedicated memory hierarchies have been designed to balance the low-cost storage provided by off-chip DRAMs and the energy-efficient access offered by on-chip SRAMs~\cite{chen2016eyeriss}. This trend has led to an increase in the use of larger on-chip memory in cutting-edge DNN accelerators. For instance, SRAM accounts for 79.2\% of the chip area and 42.5\% of the power consumption in Eyeriss (as shown in Fig.~\ref{fig:srambreakdown}.(a)), 67\% of the chip area in chiplet designs like Simba~\cite{shao2019simba}, and the latest wafer-scale chips house up to 18 GB of on-chip memory~\cite{celebras}. Thus, we see that the use of on-chip SRAM memories results in higher power \& area requirements.

\begin{table}[h!]
    \centering
    \caption{Embedded Random-Access-Memory (eRAM) \\Comparison at 65nm CMOS technology}
    \label{tab:compare_eram}
    \resizebox{\columnwidth}{!}{
    \begin{tabular}{|l|c|c|c|c|c|}
    \hline
         \multirow{1}{*}{eRAM types} & SRAM & \makecell{eDRAM\\(1T1C)} & \makecell{Symmetric \\eDRAM(3T)} & \makecell{Asymmetric \\eDRAM(2T)}\\ \hline
         Cell Size & 1$\times$ & 0.22$\times$ & 0.47$\times$ & 0.48$\times$  \\ \hline
         Avg. Static Power & 1$\times$ & 0.2$\times$ & 0.48$\times$ & 0.19$\times$  \\ \hline
         Refresh & No Ref. & Low Freq. &  High Freq. & High Freq.  \\ \hline
         Leakage & High & Low &  Low & Low  \\ \hline
         \makecell{Additional Material} & No & Yes & No & No \\ \hline
    \end{tabular}}
\end{table}

6T SRAMs have long been the preferred embedded memory choice because of their logic-compatible bit-cell, quick differential read, and static data retention \cite{khan2010trends}. However, their relatively large cell size and competing requirements for reading and writing at low operating voltages make scaling 6T SRAMs difficult in advanced CMOS technologies \cite{khan2010trends}. Recently, nonvolatile memories have captured the research community's interest due to their small cell size, low cell leakage, and fast read access operation. Earlier studies~\cite{long2019design,reis2018computing} have attempted to replace on-chip SRAM with nonvolatile memories like ReRAM, FeFET, and others. Nonetheless, the write operation in a nonvolatile memory is slower and consumes higher energy than the read operation \cite{inci2021deepnvm++,mittal2015survey, yu2021compute}. 
This can negatively impact the performance of AI chips in DNN applications, such as on-chip learning, where both on-chip read and write operations are imperative \cite{peng2020dnn+}. Another alternative to on-chip SRAM is embedded dynamic random-access memory (eDRAM). Table~\ref{tab:compare_eram} presents comparisons across different embedded memories on the same 65 nm low power CMOS process~\cite{chun2011667}. We find that 1T1C eDRAM (1 transistor and 1 capacitor) offers 4.5$\times$ higher bit-cell density and 5.0$\times$ lower static power dissipation than 6T SRAMs, even including refresh power. This results in a smaller chip size, faster memory access and increased memory density, which are the most effective methods to enhance the microprocessor performance within given power constraints. However, nonvolatile memory and conventional eDRAM (1T1C) entail a complex fabrication process, as they require specialized materials for wafer deployment~\cite{ali2019memory}. 

3T (three-transistor) and 2T (two-transistor) CMOS eDRAM gain cell designs are embedded dynamic random-access memory circuits that utilize fewer transistors per memory cell than traditional SRAM. This results in increased density and smaller area. 3T/2T eDRAM cells are made using logic devices, enabling their construction in standard CMOS processes with minimal modifications. Industrial designs have showcased that three transistors can be used to achieve approximately 2$\times$ higher bit-cell density than SRAM.
To that end, the eDRAM gain cell (3T and 2T) can reduce the on-chip SRAM area without altering the fabrication technology. As shown in Table~\ref{tab:compare_eram}, the eDRAM gain cell provides both area and energy advantages compared to on-chip SRAM. Specifically, the 2T eDRAM offers a 5.26$\times$ reduction in static power disspitation compared to SRAM. However, the use of eDRAM gain cells results in significant refresh power consumption due to smaller retention times, thereby limiting the power advantages of the eDRAM gain cells with respect to on-chip SRAM. Consequently, implementing the eDRAM gain cells in AI chips remains a viable consideration.

In deep learning applications, INT8 has emerged as the ideal numerical representation, maintaining accuracy across a wide range of tasks~\cite{jouppi2021ten}. In the 8-bit integer format, a standard for DNN quantization, errors occurring in the Most Significant Bits (MSBs) carry more weight than those in the Least Significant Bits (LSBs)~\cite{li2017understanding}. As per observations from \cite{nguyen2021zem}, the 8-bit integer data from quantized DNN tend to cluster around zero. For such small integers near zero, the MSBs are usually zeros for positive values and ones for negative values. This pattern offers a chance to increase the number of ones in positive integers by bit flipping, thus creating a dominance of ones in DNN data. The LSBs, more populated by zero bits, can bear errors with minimal effect on the final accuracy owing to their lesser significance. A recent study presents an asymmetric DNN data-encoder that boosts the frequency of bit-0\cite{nguyen2021zem} in the INT8 representations while preserving DNN performance. This idea can be further exploited in conjunction with on-chip data storage using 2T eDRAM that demonstrates an asymmetry in data retention between bit-1 and bit-0, where bit-1 offers reduced static and access energy compared to bit-0 \cite{chun2011667}. Integrating a hybrid 6T SRAM/2T eDRAM design with a one-enhancement data-encoder (that enhances the prevalence of bit-1 in INT8 representations) can optimize for both area and energy consumption on-chip. Thus, we introduce MCAIMem, a mixed memory cell based on SRAM and asymmetric eDRAM designed for area and energy-efficient on-chip AI memory. MCAIMem is adaptable, capable of accommodating various memory capacities and performance needs, making it appropriate for a broad spectrum of AI applications, from compact edge devices to extensive data centers. Our contributions are as follows:
\vspace{-0.5em}
\begin{itemize}
    \item To the best of our knowledge, we propose the first mixed 6T SRAM and 2T eDRAM cells for on-chip AI memory. We modify 2T eDRAM cells to align with SRAM cells and enhance capacity for longer retention times. Our mixed memory cells significantly reduce on-chip memory footprint for AI accelerators.
    
    \item We propose the common voltage sense amplifier (CVSA), which can be used for both SRAM and 2T eDRAM cells. By controlling the reference voltage of CVSA, we can extend the refresh period of 2T eDRAM, reducing MCAIMem's dynamic refresh energy.
    
    \item We exploit the asymmetric 2T eDRAM, where storing bit-1 consumes less energy than bit-0. Combining a one-enhancement encoder/decoder for DNN data addresses eDRAM reliability issues, such as refresh rate and retention time, significantly reducing MCAIMem's static power.
    
    \item Our MCAIMem reduces the area consumption by 48\% and improves  energy efficiency by 3.4$\times$ in on-chip AI memory systems, as demonstrated in Fig.~\ref{fig:srambreakdown}.(b), by blending the strengths of 6T SRAM and 2T eDRAM to create a high-performance, energy-efficient, and compact hybrid memory solution.

\end{itemize}

The structure of this paper is organized as follows. Section~\ref{sec:background} provides background information on gain cell eDRAM, peripheral circuitry, and the use of two's complement in DNN data representation. Section~\ref{sec:ourMCAIMem} details the comprehensive AI memory design and the operational mechanisms of AI memory. Section~\ref{sec:aiundermcaimem} discusses the impact of MCAIMem on AI applications. Section~\ref{sec:evaluation} presents hardware evaluation results from a 45 nm process technology, encompassing both circuit and system levels. Section~\ref{sec:relatedworks} examines state-of-the-art works incorporating eDRAM in AI applications. Finally, conclusions are drawn in Section~\ref{sec:conlusion}.

\section{Background}
\label{sec:background}

This section presents the background of 2T/3T eDRAM gain cell circuit designs with full CMOS technology and operation, reviews the two's complement representation in DNNs, and summarizes the challenges and requirements for designing mixed SRAM and eDRAM cell memory for AI chips.

\subsection{Embedded DRAM cell and Sensing design}
\label{sec:eDRAMbackground}
\subsubsection{3T and 2T eDRAM}

Using fewer transistors per memory cell than traditional SRAM, 3T and 2T eDRAM designs permit a smaller area, higher density, and roughly 2$\times$ greater bit-cell density. As evidenced in recent studies~\cite{choi2015refresh,giterman2018800,giterman20201,amat2018modem}, the eDRAM gain cell is currently under active development, with the newest implementations seen in 7-10nm FinFET technology. Given the typically slow rate of technological scaling, the eDRAM gain cell retains its significance and utility, as illustrated in Fig.~\ref{fig:srambreakdown}.c. In sleep mode, eDRAM cells can exhibit lower cell leakage current than SRAMs, leading to reduced static power dissipation, encompassing both leakage and refresh power components. The cell write-margin of eDRAM cells is superior to SRAMs, as there is no contention between the access device and cross-coupled latch in a gain cell. However, conventional gain cells face short retention times due to the small storage capacitor and leakage currents that exponentially vary under Process-Voltage-Temperature (PVT) variations, causing higher refresh power dissipation and/or smaller read current. The former results from the more frequent refresh operation, while the latter is due to the faster cell voltage loss.

\begin{figure}[!t]
\centering
\includegraphics[width=\linewidth]{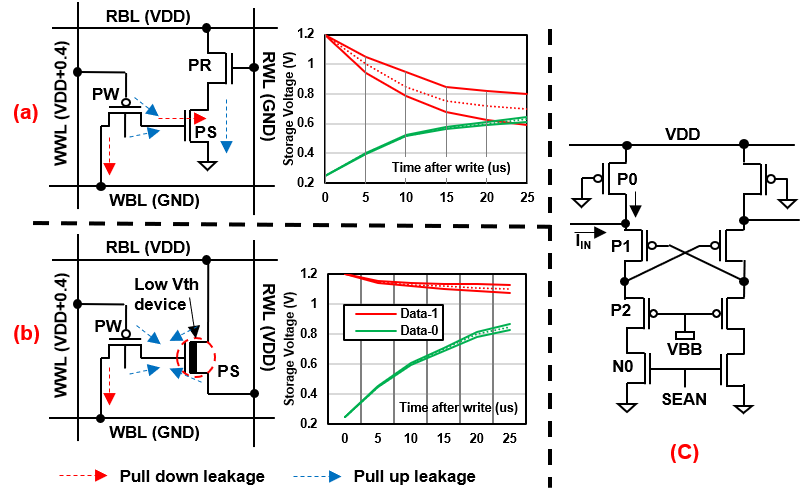}
\caption{The design of 3T and 2T eDRAM gain cells and data retention time measurement with low power CMOS 45nm. (a) the cell design and data retention time of 3T eDRAM, (b) the cell design and data retention time of 2T eDRAM, and c) The current sense amplifier of 2T eDRAM}
\label{fig:3Tand2T}
\vspace{-2em}
\end{figure}

To comprehend eDRAM gain cells, consider a conventional 3T gain cell's basic retention characteristics. In the 3T NMOS cell~\cite{chun20113t} shown in Fig.~\ref{fig:3Tand2T}.(a), PW represents the write access device, PS the storage device, and PR the read access device. In 3T eDRAM, the gate capacitor of PS is used to store the charge voltage for bit-1 or bit-0 representation. PS provides a smaller capacity compared to 1T1C eDRAM. As a result, 3T gain cells feature a decoupled read and write structure with separate Read Word-Line (RWL) and Read Bit-Line (RBL) for read access, and Write Word-Line (WWL) and Write Bit-Line (WBL) for write access. This leads to enhanced read and write-margins and flexibility in bit-cell design, allowing gain cells to scale well in future technology nodes. PW and PR are deactivated during data retention mode, leaving the storage node floating. The surrounding devices' sub-threshold, gate, and junction leakages cause the floating voltage to change over time. Data retention time relies on the aggregated leakage current entering the storage node. Monte-Carlo simulations in SPICE, representing cell-to-cell variation in a 1 Mb memory macro using low power CMOS 45nm technology, display retention time variations as illustrated in Fig.~\ref{fig:3Tand2T}.(a). With a read reference bias level of 0.65 V, both bit-1 voltage and bit-0 voltage approach the read reference bias level at the same retention time.

The innovative 2T gain cell design~\cite{chun2011667} employs fewer transistors, featuring a single high-drive current NMOS read device driven by RWL and a PMOS write device that maintains the critical bit-1 voltage near VDD. Fig.~\ref{fig:3Tand2T}.(b) presents the 2T eDRAM cell, which has a substantially different structure and operational principle compared to its 3T counterpart. The prior 3T eDRAM cell used a PMOS device as the write access transistor to extend cell retention time by counteracting NMOS gate leakage with PMOS gate overlap and junction leakages. Nonetheless, under PVT variations, leakage compensation proves insufficient as the NMOS storage device's inverted channel gate leakage dominates for bit-1, as illustrated in Fig.~\ref{fig:3Tand2T}.(a).

In the 2T eDRAM cell shown in Fig.~\ref{fig:3Tand2T}.(b), the read access transistor is substituted by the RWL signal, with a pre-charge level set at VDD. The storage transistor remains essentially off, making its gate leakage insignificant. Due to the absence of sub-threshold leakage through the read path, a low Vth transistor is proposed to improve read speed further. The stored cell exhibits asymmetry with a 0.65V read reference bias, with bit-1 unaltered while bit-0 is prone to flipping to bit-1. The balanced P and N diffusion densities are another benefit of the proposed 2T asymmetric cell. This paper aims to leverage this characteristic to minimize both static and dynamic energy consumption since bit-0 requires more energy to flip.

\subsubsection{2T eDRAM Sense Amplifier}

In gain cells, the NMOS gate capacitor is used to store charge, rendering them sensitive to voltage changes. Directly accessing NMOS could cause a stored bit to flip. As a result, conventional gain cells need a current sense amplifier to sense the storage node. For 2T cell designs, the RBL must exhibit a limited swing to avoid read failures caused by unselected cells' leakage current. However, a smaller voltage swing results in a poorer read sensing margin. The asymmetric 2T gain cell further complicates the situation by utilizing a low Vth read device to achieve faster read speeds, while keeping the speed-critical bit-1 voltage close to VDD. To tackle this issue, a Current-mode Sense Amplifier (C-S/A) is employed in 2T eDRAM design, maintaining the RBL voltage near VDD during sensing and permitting the connection of multiple low Vth cells to a single RBL.

The C-S/A, depicted in Fig.~\ref{fig:3Tand2T}.(c), consists of a cross-coupled PMOS latch (P1) and a pseudo-PMOS diode (P2) driven by the negative supply VBB, which is easily accessible on the chip for WWL under-driving. A negative WWL is essential for a PMOS device to write a bit-1 into the cell without incurring a threshold voltage loss. Both PMOS pairs (P1 and P2) operate in saturation mode, providing improved matching. However, this C-S/A design is exclusively used for reading the storage bit in 2T eDRAM cells, as bit-0 still necessitates periodic write-back to avert bit flipping. Hence, an extra write circuit is required for the write operation, leading to inefficiency due to the small size of the 2T eDRAM cell and the substantial overhead needed for read/write circuits. In this study, we propose a modification to the 2T eDRAM cell that enhances its capacity and aligns it with the SRAM cell size. This slight alteration in the 2T eDRAM cell's design simplifies the sensing circuit, making it compatible with both SRAM and 2T eDRAM. Unlike the conventional 2T which only leverages the asymmetric characteristics of 2T eDRAM for sensing in a small voltage swing with C-S/A, our approach extends the reading process due to voltage sensing. The storage charge, in our case, has a larger voltage margin ranging from 0 to 0.8V for bit-0 and from 0.8 to 1.0V for bit-1. More in-depth discussion is presented in Section~\ref{sec:mixedcelldesign}.



\subsection{Two's complement and one-enhancement method in DNN representation}
\label{sec:twocomplement}

In deep neural networks (DNNs), the choice of data representation significantly impacts accuracy, computational complexity, and power consumption. Two's complement representation is a popular format for signed integer values in DNNs because it simplifies arithmetic operations, particularly multiplication and addition. This format represents negative numbers as the two's complement of the corresponding positive number's binary representation, easing hardware implementation and reducing circuit complexity.

At present, INT8 is regarded as the optimal representation for DNN inference, maintaining accurate results~\cite{jouppi2021ten}. The 8-bit two's complement quantization, using approaches such as~\cite{choi2018pact}, is extensively adopted and outperforms other quantization techniques. In this work, we opt for the 8-bit two's complement as the benchmark for designing the on-chip buffer.

\begin{figure}[!t]
\centering
\includegraphics[width=0.8\linewidth]{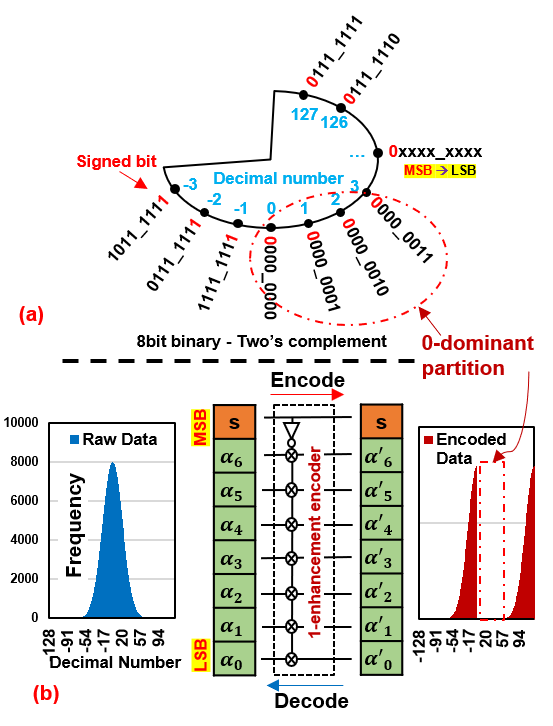}
\caption{(a) Illustration of an 8-bit integer represented in two's complement form, showcasing the binary encoding of both positive and negative integers. (b) Demonstrating the impact of the One Enhancement Encoder/Decoder, where the original bit data is converted to 1-bit data based on its sign bit. This figure also presents measurements related to the weight of a ResNet-50 neural network, highlighting the applied encoding technique’s implications on data representation and subsequent computations.}
\label{fig:twocomplement}
\vspace{-1em}
\end{figure}

As depicted in Fig.~\ref{fig:twocomplement}.(a), the first bit, known as the signed bit, determines whether the number is positive or negative. As noted in ZEM\cite{nguyen2021zem}, DNN data typically falls within a narrow range (e.g., [-50, 50]). Negative values near zero possess 1-dominant bits, while the corresponding positive numbers exhibit 0-dominant bits. Converting 0-dominant bits involves flipping all data bits based on the signed bit. As shown in Fig.~\ref{fig:twocomplement}.(b), constructing the encoder requires only one INV and seven XOR gates, transforming raw data into 1-dominant bit data. The one-enhancement encoder encodes DNN data into 1-dominant bit data. The decoder processes the encoded data by flipping the bits according to the signed bit, thereby reconstructing the original data. Our work, unlike ZEM\cite{nguyen2021zem}, aims to create 1-dominant data to decrease refresh and static energy usage in mixed-cell memory design when storing DNN data. In this paper, the DNN data stored in on-chip memory undergoes encoding and decoding before computation.

\subsection{Summary of design challenges and requirements}

As we consider the requirements of the one-enhancement encoder/decoder, the signed bit acts as the control bit, dictating when encoding or decoding operations should be executed. It is crucial to protect the signed bit from errors. In this paper, we take advantage of 2T eDRAM to enhance area and energy efficiency. Nonetheless, 2T eDRAM demands more frequent refresh operations due to the accelerated cell voltage loss. To secure the signed bit, we allocate it to 6T SRAM and map the remaining bits to 2T eDRAM. As a result, we need to address the following challenges when designing mixed SRAM and eDRAM cells:

\begin{itemize}
\item \textit{Compatibility}: Ensure seamless integration of SRAM and eDRAM cells within a single memory architecture, preserving compatibility with existing manufacturing processes.
\item \textit{Density and Area Efficiency}: Achieve high memory density and area efficiency without sacrificing performance or increasing chip complexity.
\item \textit{Retention Time and Refresh Rate}: Address the intrinsic retention time limitations of eDRAM cells and optimize refresh rates to minimize power consumption without compromising data integrity.
\item \textit{Scalability}: Develop memory architectures that can be effortlessly scaled to accommodate diverse memory capacities and performance requirements for various AI applications, spanning from edge devices to data centers.
\item \textit{Reliability}: Guarantee the robustness and reliability of the mixed memory design under different operating conditions, especially in AI workloads involving frequent read and write operations.
\end{itemize}

In this paper, we will tackle the abovementioned concerns and introduce an efficient on-chip memory design suitable for AI applications. Further details will be provided in the subsequent section.

\section{Our MCAIMem}
\label{sec:ourMCAIMem}

\begin{figure}[!t]
\centering
\includegraphics[width=0.8\linewidth]{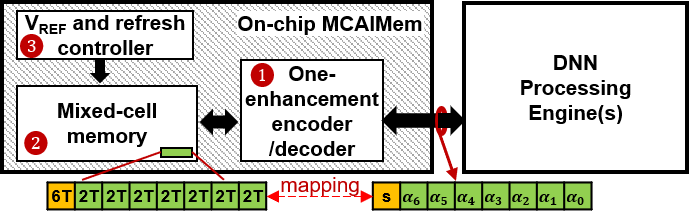}
\caption{Overall architecture of MCAIMem. It comprises 1) One-enhancement encoder/decoder, 2) Mixed-cell memory, and 3) Reference voltage and refresh controller}
\label{fig:overviews}
\vspace{-2em}
\end{figure}

This section introduces MCAIMem, our innovative on-chip mixed-cell memory design specifically developed for AI chips. As illustrated in Fig.~\ref{fig:overviews}, MCAIMem consists of three key components: 1) the mixed SRAM/eDRAM cell memory, including mapping schemes and circuit-related designs, 2) the one-enhancement encoder/decoder, and 3) the reference voltage controller responsible for lengthening the refresh operation duration. The on-chip MCAIMem is a buffer for AI accelerators, storing both weights and activations during computation. Data transferred from off-chip DRAM is retained in MCAIMem and subsequently employed by the DNN processing engine, which can range from traditional CPUs/GPUs to systolic arrays or computing-in-memory (CIM) architectures.

Inbound/outbound data must pass through the one-enhancement encoder module for encoding/decoding, which will be discussed in Section~\ref{sec:oneenhancement}. The encoded data is preserved in our mixed-cell memory design, a combination of 6T SRAM and 2T eDRAM, aimed at minimizing the area footprint of on-chip memory. Further details about the mixed-cell memory design will be provided in Section~\ref{sec:mixedcelldesign}. Owing to the inclusion of 2T eDRAM, periodic refresh operations are necessary. The refresh controller will be discussed in Section~\ref{sec:vrefreshcontroller}.

\subsection{One-enhancement encoder/decoder and data mapping}
\subsubsection{One-enhancement encoder/decoder module}
\label{sec:oneenhancement}

To store data within our on-chip MCAIMem, it is necessary to first encode it using the One-enhancement encoder/decoder module. After conducting synthesis on 45nm technology node, we performed an experimental assessment of the one-enhancement encoding/decoding module. The power consumption of the module, measured at 1.35$\times10^{-2}$mW, constitutes a mere 0.007\% of the total memory power, rendering its impact negligible. In terms of area, this module occupies only 35.2$um^2$, which is a trifling 0.004\% when compared to the 108KB memory size. These metrics underscore the module's inconsequential influence on both power usage and spatial requirements, especially when juxtaposed with the vast expanse of memory cells. Moreover, the delay associated with the encoder/decoder stands at just 0.23ns. Even with a clock period of 1ns (corresponding to a clock frequency of 1GHz), there's a comfortable slack of 0.67ns, ensuring the absence of timing violations. Consequently, the encoder/decoder's delay poses no threat to the system's overall performance.

As discussed in Section~\ref{sec:twocomplement}, incoming data undergoes flipping based on its signed bit before being stored in our mixed-cell memory. By enhancing the raw bit data to predominantly 1-bit values, the overall energy consumption of the memory can be reduced, as the cells are optimized to store and access 1-bits more efficiently. The signed bit from the two's complement representation can be utilized to perform this enhancement, as it is either 0 or 1 depending on the sign of the number. The encoder modifies the input data such that more 1-bits are present, while the decoder reverses this process to recover the original data.A significant portion of DNN data is either 0 or values near 0. For instance, pruning results in 20-80\% of the data being 0~\cite{li2016pruning}. Since a majority of the data is close to zero, enhancing the representation to produce more bit-1 values can lead to power efficiency in memory systems without sacrificing data integrity or accuracy.

\begin{figure}[!t]
\centering
\includegraphics[width=0.9\linewidth]{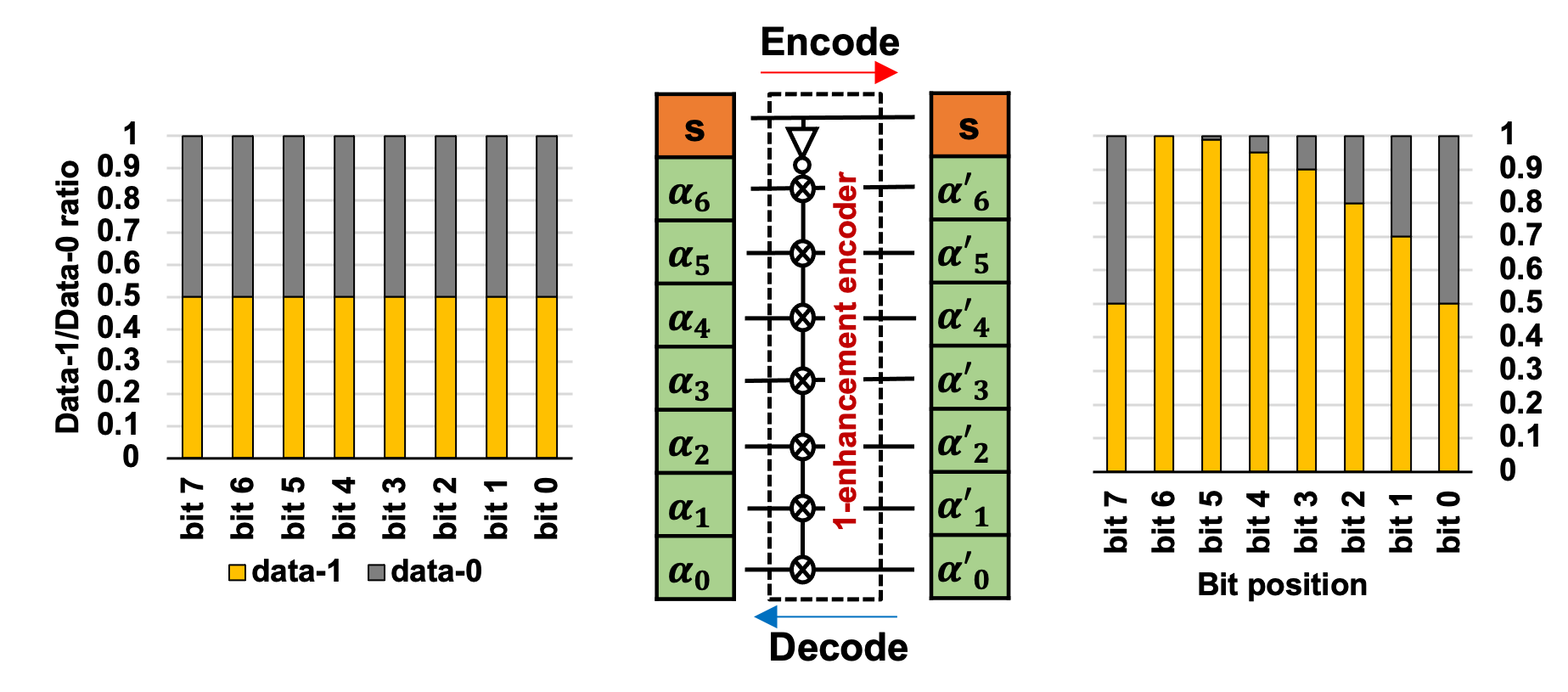}
\caption{Visual representation of the bit statistic histogram for ResNet50's weights pre- and post-function of the One-Enhancement encoder/decoder. The initial data is assumed to have an equal probability of 50\% for both 0 and 1 bits, demonstrating the transformation and distribution of bit values upon the encoding and decoding process.}
\label{fig:oneenhance}
\end{figure}

As illustrated in Fig.~\ref{fig:oneenhance}, the 6th, 5th, and 4th bits mostly convert to bit-1, making it highly efficient to map them to 2T eDRAM cells. The 0th, 1st, 2nd, and 3rd bits continue to contain a considerable number of bit-0s. When using 2T eDRAM to store such bits, retention errors might occur. However, due to the asymmetric nature of 2T eDRAM, only flipping errors from 0 to 1 are considered retention errors. These errors can potentially impact the output of DNN applications. To ensure these errors do not adversely affect DNN application outcomes, we will evaluate their implications in Section~\ref{sec:retentionaffectAIoutcome}.

\subsubsection{Mixed cell mapping scheme}
As discussed in Section~\ref{sec:eDRAMbackground}, using 2T eDRAM can lead to errors resulting from its short retention time. These errors may arise when storing DNN data in our MCAIMem. Based on the one-enhancement encoder/decoder, the control bit is of utmost importance since an error in the control bit would cause errors in all the remaining bits. Consequently, we need to ensure the control bit is well protected when storing DNN data in 2T eDRAM, while allowing for approximation in the remaining bits. The proposed bit mapping should be as follows: 1) map the control bit to 6T SRAM, and 2) map the 7 least significant bits (LSBs) to 2T eDRAM.

\begin{figure}[!t]
\centering
\includegraphics[width=0.5\linewidth]{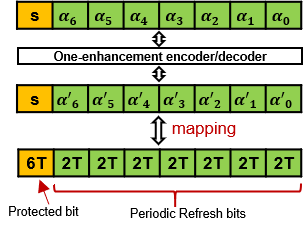}
\caption{This diagram illustrates our mapping method, where the signed bit is allocated to 6T SRAM, while the remaining bits are designated to 2T eDRAM. This approach effectively distributes data between the two memory types, optimizing their respective storage capabilities.}
\label{fig:mapping}
\end{figure}

As illustrated in Fig.~\ref{fig:mapping}, one 6T SRAM cell is allocated for the signed/control bit, while the following seven bits are mapped to 2T eDRAM cells. Incoming DNN data is first encoded by the One-enhancement encoder and then stored in the mixed-cell array. The signed bit/control bit is securely stored in the 6T SRAM, while the remaining bits are flipped according to the signed bit and stored in the 2T eDRAM, as shown in Fig.~\ref{fig:mapping}.

This memory mapping approach ensures the safety of the signed/control bit in the 6T SRAM, while requiring periodic refresh operations for the remaining bits to prevent data flipping. This mechanism is crucial for maintaining the dominance of bit-1 in the majority (around 80\%) of DNN data. Due to the characteristics of 2T eDRAM, storing bit-1 consumes less energy than storing bit-0. Therefore, static energy savings can be achieved by using the one-enhancement encoder in conjunction with asymmetric 2T eDRAM.

\subsection{Mixed-Cell Memory design}
\label{sec:mixedcelldesign}

As mentioned in the mapping method, our mixed-cell memory design consists of one 6T SRAM cell and seven 2T eDRAM cells. To integrate these cells, we need to modify the circuit designs of both 6T SRAM and 2T eDRAM. In this work, we propose modifications to the 2T eDRAM and minor changes to the 6T SRAM. Additionally, we suggest a voltage sense amplifier that can be used for both 6T SRAM and 2T eDRAM. The details of these adjustments will be discussed in the following subsections.

\subsubsection{Enhance the retention time of asymmetric 2T eDRAM cell}
\label{sec:enhance2TeDRAM}

\begin{figure}[!t]
\centering
\includegraphics[width=0.7\linewidth]{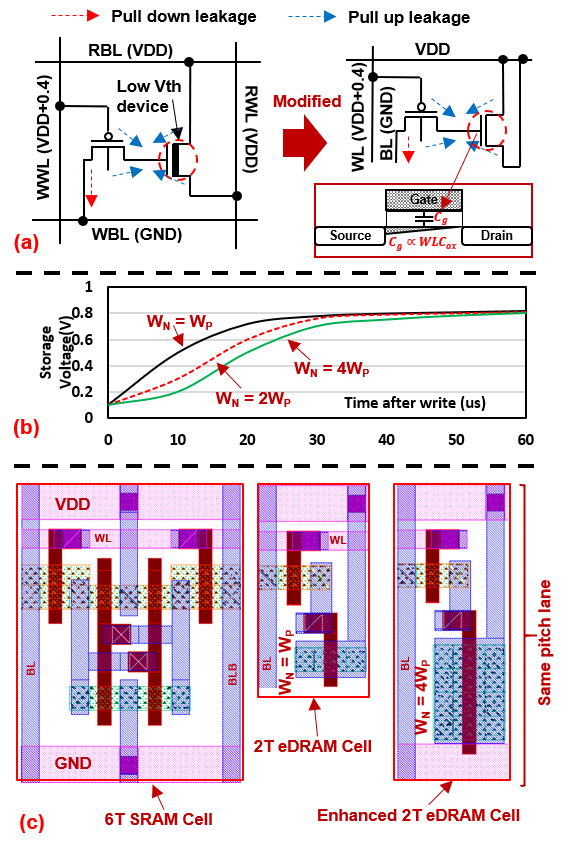}
\caption{(a) The modified design of 2T eDRAM in which the Read Bit Line (RBL) and Read Word Line (RWL) are directly connected to VDD, eliminating the need for superfluous threshold voltage (Vth) NMOS transistors. (b) Discussion on the retention time and the influence of the storage node width in 2T eDRAM. (c) Depiction of the layout design for the 6T SRAM, traditional 2T eDRAM, and our proposed design featuring a stretched-width 2T eDRAM.}
\label{fig:storagenode}
\vspace{-2em}
\end{figure}

In combining the designs of 6T SRAM and 2T eDRAM, we encounter challenges regarding pitch lane mismatches and the need for a mixed sense amplifier suitable for both 6T SRAM and 2T eDRAM. This is due to the considerably smaller size of the 2T eDRAM compared to the 6T SRAM. To resolve the pitch lane issue, we adjust the size of the 2T eDRAM's storage transistor. As depicted in Fig.~\ref{fig:storagenode}.(c), pitch lane mismatches can occur when designing the cell layout for 6T SRAM and 2T eDRAM. The 2T eDRAM cell occupies only 60\% of the area compared to an SRAM cell. Therefore, we can increase the width of the 2T eDRAM up to 4$\times$ to align with the design of the 6T SRAM cell.

As discussed in section~\ref{sec:eDRAMbackground}, the 2T eDRAM consists of two primary components: the access transistor and the storage node. The gate capacitor ($C_g$) in the storage node stores the charge voltage representing bit-1 or bit-0, as shown in Fig.~\ref{fig:storagenode}.(a). The capacity of the NMOS gate is defined as $Cg \propto WLC_{ox}$. By increasing the width of the NMOS gate, we not only enhance the storage node's capacity but also improve the 2T eDRAM's retention time. Fig.~\ref{fig:storagenode}.(b) demonstrates the SPICE simulation of eDRAM design using CMOS 45 nm technology; when storing bit-0, the retention time is significantly extended. For example, when the storage node width is increased by four times, the time required for the charge to change from 0.18V to 0.8V doubles.

Moreover, increasing the storage node's capacity offers additional benefits. It makes the 2T eDRAM more robust against read-disturb effects. This allows us to remove the RWL and RBL in the 2T eDRAM and connect them directly to VDD. The NMOS transistor now functions solely as a capacitor for storage. The gate leakage from VDD of the storage node, along with the gate leakage and junction leakage of the write transistor, refills the storage node's charge to bit-1. As a result, the asymmetric characteristic of 2T eDRAM remains intact. With this design, we expect to store bit-1 without retention time, while bit-0 will need periodic refresh operations to maintain their discharged state. Thus, we can directly connect the drain and source of the storage node to VDD, as shown in Fig.~\ref{fig:storagenode}.(a). In this study, we utilize the pull-up leakage current to sustain the '1' bit and to store the '0' bit, thereby eliminating the need for changes in doping and gate oxide thickness typically required for Low Voltage Threshold (LVT) devices. This approach renders such modifications unnecessary.

\subsubsection{The adaptation of SRAM cell design in mixed memory cells}
While we made significant modifications to the storage node of the 2T eDRAM, we retained the PMOS access transistor from the conventional 2T eDRAM design. This decision was made to ensure that only bit-0 changes while bit-1 remains unchanged. To minimize the subthreshold leakage of the PMOS and ensure that the pull-down leakage path is always lower than the pull-up leakage path, we applied VDD+0.4V. However, using PMOS transistors as the access transistor might conflict with the design of the access transistor in 6T SRAM.

\begin{figure}[!t]
\centering
\includegraphics[width=0.7\linewidth]{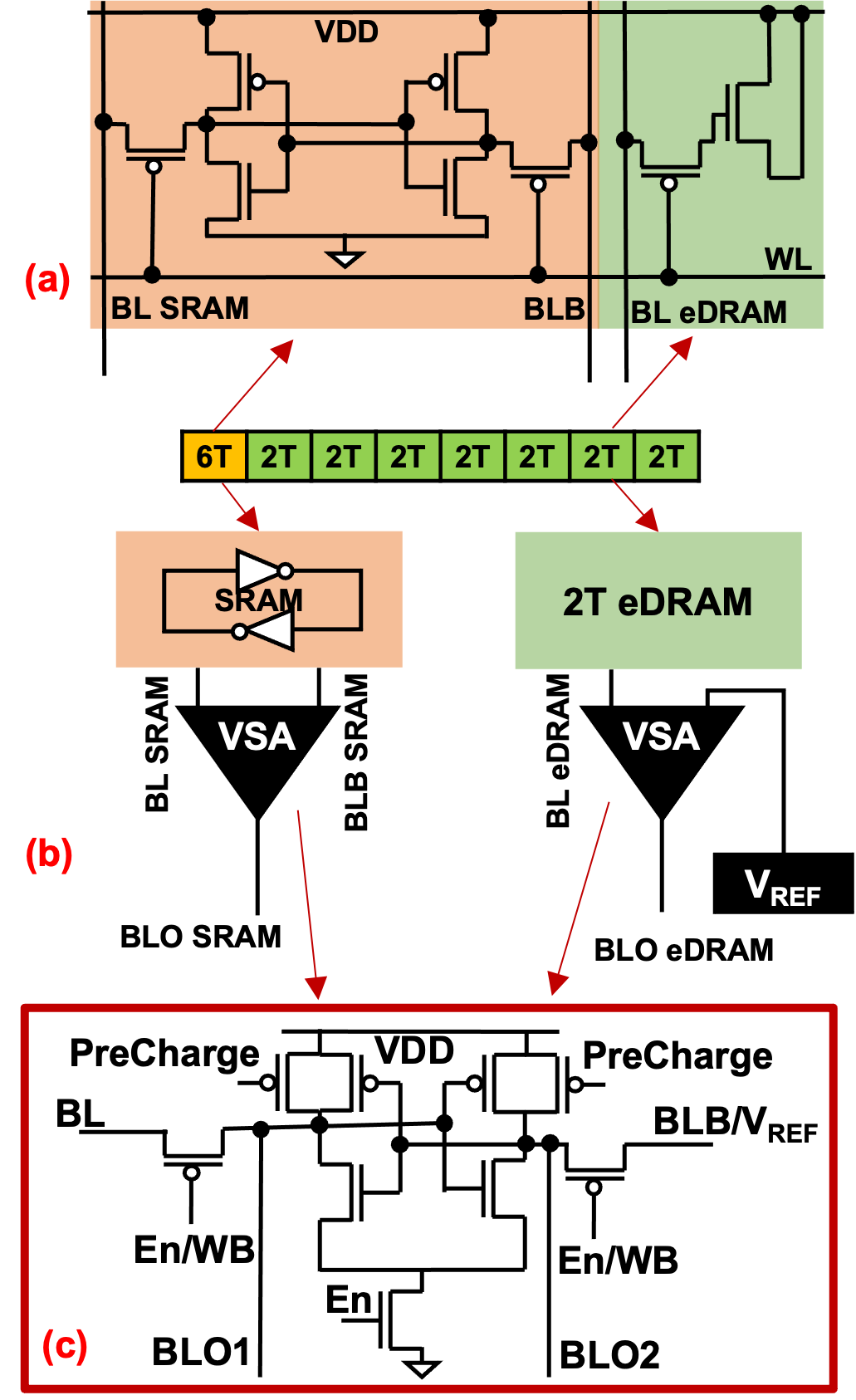}
\caption{(a) 6T SRAM cell and 2T eDRAM cell design, (b) the connectivity between 6T SRAM and 2T eDRAM with the multiple purpose sense amplifier, and (c) the common sense amplifier circuit design}
\label{fig:mixedcircuit}
\vspace{-2em}
\end{figure}

To address this issue, we made minor modifications to the SRAM cell design as shown in Fig.~\ref{fig:mixedcircuit}.(a), changing the access transistor in the SRAM to PMOS as well. By aligning the access transistor types in both memory cells, we facilitate the integration of 6T SRAM and 2T eDRAM designs while maintaining the desired functionality and performance.


\begin{figure}[!t]
\centering
\includegraphics[width=\linewidth]{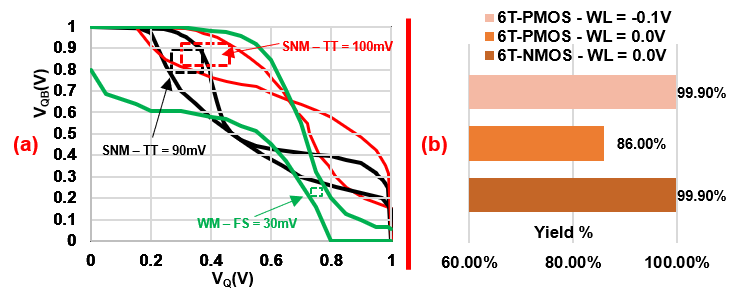}
\vspace{-2em}
\caption{(a) Examines the readability and writability of 6T SRAM with various access transistor configurations. Black, red, and green lines represent the read static noise margin (SNM) of 6T SRAM with NMOS and PMOS access transistors, and the worst write margin with PMOS access transistor, respectively. (b) Compares the write yield of 6T SRAM with different access transistor configurations by conducting a Monte Carlo simulation 1000 times at 25$^o$C.}
\label{fig:yield}
\vspace{-2em}
\end{figure}
By modifying the access transistor in a 6T SRAM bit cell (refer to Fig.\ref{fig:yield}.a), we observe a higher read static noise margin (SNM) of 100mV with a pMOS access transistor (red line) compared to 90mV with an nMOS transistor (black line). However, pMOS transistors have lower writeability. As node QB discharges and Vgs decreases, the transistor weakens and shuts off when QB dips below the threshold voltage, leading to a constrained write margin of 30mV at FS corner (green line). This issue can be mitigated by applying a small negative voltage during the 6T SRAM write process\cite{nabavi2017290}. With a -0.1V on the word line (WL), the write yield of the pMOS access transistor increases to match that of an nMOS transistor as shown in Fig.\ref{fig:yield}.b.

\subsubsection{Voltage sense amplifier circuit for both SRAM and 2T eDRAM}

As discussed in section~\ref{sec:eDRAMbackground}, the 2T eDRAM decouples the read and write paths, which means that separate circuits are required for read and write operations. Additionally, the short retention time of 2T eDRAM necessitates periodic refresh operations to maintain the data. One of the major challenges in designing mixed cells is providing a mixed sense amplifier that caters to both 6T SRAM and 2T eDRAM.

In 2T eDRAM, a current sense amplifier is needed to detect small gains in the read path without disturbing the data in the storage node. However, the refresh operation demands that the read data be written back to the storage node, resulting in substantial peripheral circuit overhead. As mentioned in section~\ref{sec:enhance2TeDRAM}, by increasing the width size of the 2T eDRAM by 4$\times$, the design can resist read-disturb. Therefore, we propose a voltage sense amplifier, as shown in Fig.~\ref{fig:mixedcircuit}.(c), for both 6T SRAM and 2T eDRAM. This enables read and write operations for both 2T eDRAM and 6T SRAM to be identical. The primary distinction between 6T SRAM and 2T eDRAM when connected to the voltage sense amplifier is that both the BL and BLB of 6T SRAM are connected. In contrast, for 2T eDRAM, only one BL connects to the sense amplifier, while the BLB of the sense amplifier links to the reference voltage ($V_{REF}$), as illustrated in Fig.~\ref{fig:mixedcircuit}.(b).

Not only does this simplify read and write operations, but the use of a voltage sense amplifier also enables writing data back to the 2T eDRAM storage node during read operations. This streamlines the refresh process, as only one read operation is needed to perform the refresh, rather than the conventional sequence of read and write back operations in standard 2T eDRAM designs.

\subsubsection{The read and write operation of the voltage sense amplifier}

\begin{figure}[!t]
\centering
\includegraphics[width=0.9\linewidth]{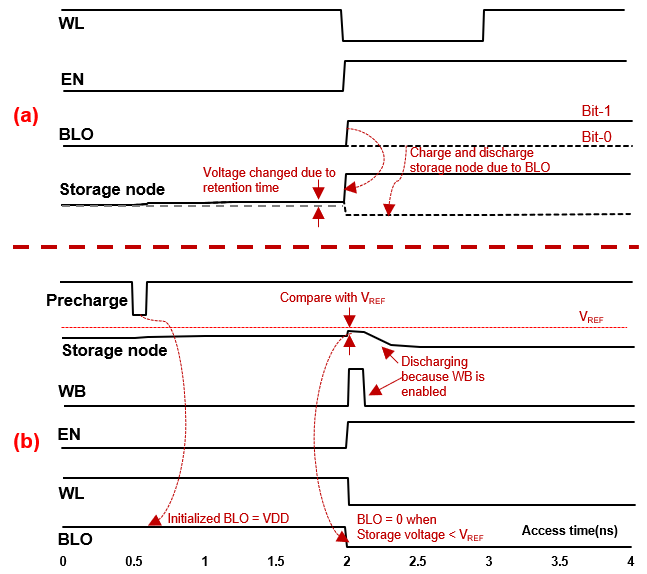}
\caption{Illustrating the sequence of write and read operations in MCAIMem. (a) the write sequence, and (b) the read sequence. This sequence is applicable to both 6T SRAM and 2T eDRAM.}
\label{fig:writeandread}
\end{figure}

Fig.~\ref{fig:writeandread}.(a) demonstrates the write operation using the voltage sense amplifier (VSA). The process begins by applying voltage to the bit-line out (BLO1), followed by enabling the sense amplifier through the enable signal (EN). This action causes the bit-line to charge or discharge. When the word-line (WL) is activated, the bit-line voltage alters the data within the 6T SRAM or 2T eDRAM. For the SRAM, the PMOS access transistor in the 6T SRAM must be weaker than the storage node for the write operation to be successful. In the case of the 2T eDRAM, the charge or discharge of the storage node occurs similarly to the SRAM.

Fig.~\ref{fig:writeandread}.(b) presents the read operation of the VSA. To initialize the sense amplifier for the read operation, the precharge is enabled, charging BLO1 and BLO2 to 1. For the 2T eDRAM, a reference voltage can be applied to the bit-line bar (BLB). This reference voltage ($V_{REF}$) is employed to compare the voltage in the storage node and determine the data output of BLO1. Once the WL and EN in the sense amplifier are enabled, the storage node will either recharge or discharge the bit-line. If the bit-line (BL) voltage is greater than $V_{REF}$ within the sense amplifier, BLO1 is set to 1; if the BL voltage is less than $V_{REF}$, BLO1 is set to 0. However, the read operation could potentially destroy the data stored in the 2T eDRAM. As a result, it is crucial to disable the WB in the sense amplifier to recharge the storage node in the 2T eDRAM. For the 6T SRAM, the BLB of the VSA is connected to the BLB of the SRAM cell.

\subsection{Reference Voltage and Refresh controller}
\label{sec:vrefreshcontroller}

As depicted in Fig.~\ref{fig:writeandread}.(b), our revised 2T eDRAM's read operation allows for the deactivation of the write-back (WB) signal. Due to the bitline voltage, the storage node can be charged or recharged, rendering the MCAIMem refresh operation as simple as executing a read operation.

Our mixed-cell memory design incorporates the 2T eDRAM, which requires periodic refresh operations. We opt for the standard periodic refresh method, as described in~\cite{baek2013refresh}, also known as the global refresh operation. In this approach, a refresh operation must be performed on each row of MCAIMem within 12.57us. To elaborate, the ordinary refresh cycle interval is calculated by dividing the refresh time by the number of rows. While the static power consumption for storing DNN's data is significantly reduced due to the one-enhancement encoder, bit-0 still requires frequent refresh operations to ensure the safety of the DNN's data. This module is responsible for determining the reference voltage for the 2T eDRAM sense amplifier, which aids in extending the refresh period and reducing the dynamic refresh energy for bit-0 in DNN's data. The reference voltage decision and its detailed discussion can be found in section~\ref{sec:refreshextention}.

\section{Mitigating DNN accuracy loss under MCAIMem}
\label{sec:aiundermcaimem}

In this section, we will explore the impact of MCAIMem on the outcomes of DNN applications. First, we will examine the influence of retention error on DNN performance. Second, we will discuss methods to extend the refresh period for energy-efficient DNN applications on AI chips.

\subsection{Retention error affects the AI chip outcomes}
\label{sec:retentionaffectAIoutcome}

\begin{figure}[!t]
\centering
\includegraphics[width=\linewidth]{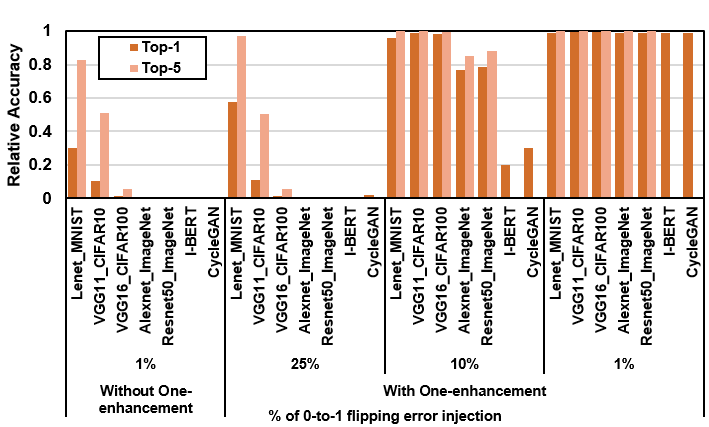}
\caption{The sensitivity of DNN accuracy under the retention error of MCAIMem with and without One-enhancement. The retention error injection range from 25\% to 1\%. Please note that the retention error injection occurs for both weight and activation.}
\label{fig:inject}
\end{figure}

DNNs are known for their robustness, with errors typically being minor and only occurring in the LSBs~\cite{li2017understanding}. In MCAIMem, our memory configuration comprises a mix of one SRAM and seven 2T eDRAM. In the mixed-cell memory design, potential errors chiefly originate from the 2T eDRAM due to retention errors, thus necessitating an exploration of their impact on DNN applications. The 2T eDRAM demonstrates a bit-0 to bit-1 flip in under 1\% of cases before 12.57$\mu$s and in over 25\% post 13$\mu$s, with no observed errors for bit-1, as illustrated in Fig.~\ref{fig:vrefchange}(b). This informs our utilization of an error injection method to evaluate the flipping error rate and discern its effect on the accuracy of DNN applications, which subsequently influences the refresh period of our MCAIMem. Notably, in this research, retention time issues are exclusive to the 2T eDRAM, not affecting the SRAM. Thus, we deliberately inject errors into both the weight and activation of DNN data before every computation, allowing the cumulative effect of the errors. We’ve devised two methods: initially, we inject errors into the DNN—absent the one-enhancement encoder/decoder—where only bit-0 is flipped at a predefined error rate. Alternately, errors are injected into bit-0 post-application of the one-enhancement encoder and prior to decoding the data, with our flipping error rate spanning from 1\% to 25\%.

We carried out simulations on several CNNs, such as LeNet, VGG11, VGG16, AlexNet, and ResNet50~\cite{lecun2015lenet,szegedy2015going,krizhevsky2012imagenet,he2016deep}, using datasets like MNIST, CIFAR10, CIFAR100, and ImageNet. In language modeling, we utilized I-BERT~\cite{kim2021bert}, the integer version of BERT, with the GLUE datasets~\cite{wang2018glue}. As for generative modeling, we used the quantized version of CycleGAN~\cite{zhu2017unpaired} with the horse2zebra dataset~\cite{wang2020gan}. This comprehensive evaluation can reveal the extent to which different retention error levels affect DNN accuracy and the effectiveness of the One-enhancement method. For each error rate specified, a comparison of DNN accuracy can be made when MCAIMem is used with and without the One-enhancement encoding. It is noteworthy that, in the case of GANs, the accuracy of the outputs can't be directly measured. Therefore, we rely on the mean relative error to quantify the difference between our GAN and the original model. This comparison will assist in evaluating how well the One-enhancement technique curtails the impact of retention errors on DNN accuracy.

Fig.~\ref{fig:inject} demonstrates that without the application of the one-enhancement encoder/decoder, the DNN accuracy plummets to zero across various networks. This drop is due to only the signed bit being safeguarded in SRAM, while the remaining bits are still susceptible to retention errors. However, upon implementing the one-enhancement encoder/decoder, a significant majority of MSBs become one bits and are not prone to flipping, while a small number of LSBs that retain zero bits may encounter retention errors. As a result, an injection error of up to 1\% can be tolerated for ImageNet on AlexNet/ResNet50, GLUE on I-BERT, and horse2zebra on CycleGAN. Furthermore, the MNIST and CIFAR10/100 datasets demonstrate a higher resilience to errors, allowing for retention errors of up to 25\%. When the one-enhancement technique is employed, significant bit errors in the MSB are introduced to the weight and activation values, which noticeably decreases the accuracy of the inference task. Nevertheless, this approach proves beneficial for extending the retention time of eDRAM. Although it introduces bit errors, they predominantly affect the LSB and thus, do not significantly impair the inference process's accuracy.

In DNN applications, the outcome is crucial for determining AI performance. Preserving the accuracy of DNN output is a more significant factor than hardware performance and energy consumption. Consequently, our MCAIMem should adhere to these requirements and consider the maximum retention error of up to 1\% in our mixed-cell design.

\subsection{Refresh period extension with an adaptive $V_{REF}$}
\label{sec:refreshextention}

\begin{figure*}[!t]
\centering
\includegraphics[width=0.8\linewidth]{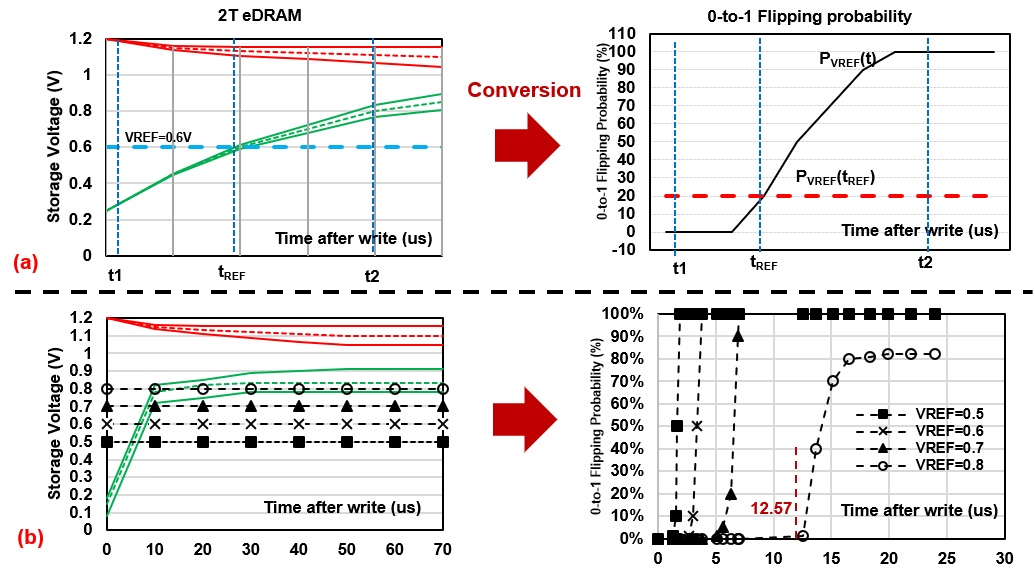}
\vspace{-1em}
\caption{(a) A statistical development method of 0-to-1 flipping error probability model of MCAIMem using the Monte Carlo simulation and $V_{REF}$, (b) 0-to-1 flipping error probability model with various $V_{REF}$ values (0.5V, 0.6V, 0.7V, and 0.8V). These simulations will provide insights into how the choice of $V_{REF}$ affects the read stability, retention time, and energy efficiency of the MCAIMem design. Analyzing the error probability results for these various reference voltages can help designers identify the optimal $V_{REF}$ value that strikes a balance between robustness, retention time, and energy efficiency for the MCAIMem memory system.}
\label{fig:vrefchange}
\vspace{-2em}
\end{figure*}

In Section~\ref{sec:eDRAMbackground}, we explained that bit-0 tends to flip to bit-1 over time, and only bit-0 is subject to retention errors. To maintain DNN output accuracy with the one-enhancement encoder, the maximum allowable retention error is 1\%. As a result, we need to determine a refresh time that preserves DNN output accuracy by developing an error model based on bit-0's retention time.

In 2T eDRAM, leakage current causes bit-0 to have a propensity to flip to bit-1 after a specific duration. This leads to variations in bit-0 readings, depending on the 2T eDRAM access time. To compute the 0-to-1 flipping probability, we execute a Monte Carlo simulation, generating a multitude of variation samples of 2T eDRAM storing bit-0. We then count the number of flipped bits concerning the total number of 2T eDRAM samples, considering the access time and specific reference voltage ($V_{REF}$), as depicted in Fig.~\ref{fig:vrefchange}.(a). This error flipping model assists in identifying the optimal $V_{REF}$ for achieving a balance between robustness, retention time, and energy efficiency in MCAIMem.

We conducted a Monte Carlo simulation 100,000 times at a temperature of 85$^o$C, which reflects typical desktop and server working conditions, where temperatures range from 25-85$^o$C~\cite{kim2014flipping}. This involved assessing data shifts in the storage node and varying the access time between 0 and 20 microseconds while simultaneously reading the data. We then compare this read-out data to the reference voltage ($V_{REF}$) to simulate the sense amplifier model, which determines whether the output data is 1 or 0. Fig.~\ref{fig:vrefchange}.(b) illustrates that with $V_{REF}$ at 0.5, a 1\% flipping probability initiates at 1.3 microseconds. Conversely, with $V_{REF}$ at 0.8, a 1\% flipping probability starts at 12.57 microseconds. The graph indicates that the flipping probability slope is steep, meaning that extending the refresh period based on a specific $V_{REF}$ yields minimal refresh power reduction. Nevertheless, adjusting $V_{REF}$ allows us to lengthen the required refresh period. As a result, we choose a $V_{REF}$ of 0.8V to maximize bit-0's refresh period and minimize dynamic refresh operations in mixed-cell memory.

\section{Evaluation}
\label{sec:evaluation}

Our study primarily concerns AI chips intended for server and desktop applications, operating within a temperature range of 25$^{o}$C to 85$^o$C. We specifically do not consider voltage variations, concentrating solely on process variations through Monte Carlo simulation. We structure our discussion into two main segments in this section. Initially, we delve into circuit simulation, followed by an examination of system simulation. The latter particularly explores DNN applications powered by our proposed MCAImem design.

\subsection{Circuit evaluation}

\begin{figure}[!t]
\centering
\includegraphics[width=\linewidth]{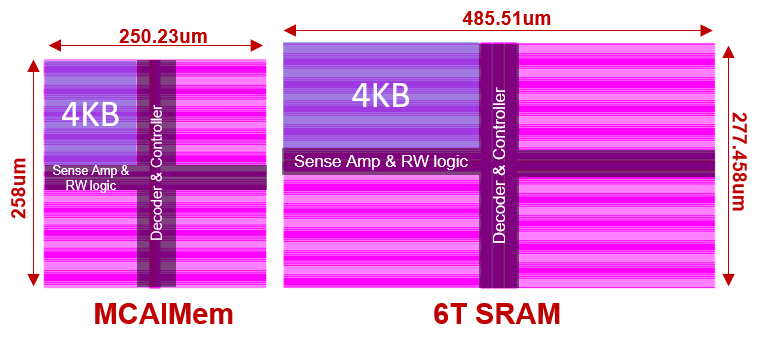}
\vspace{-2em}
\caption{Our MCAIMem's 16KB bank layout demonstrates a 48\% area reduction when compared to the equivalent 6T SRAM 16KB bank layout. Please note that 1MB memory comprises 64 banks}
\label{fig:layout}
\vspace{-1em}
\end{figure}

In the circuit evaluation, we create layouts for 1MB 6T SRAM, 2T eDRAM, and our mixed-cell memory using CMOS 45nm technology. We calculate the chip area for these embedded RAMs based on their layout sizes and compare them. Additionally, we extract the SPICE models of these memories and perform the post-simulations to analyze the static power, read, and write operations for each memory type. Table. ~\ref{tab:compare_memory}. summarizes the characterization results.

\begin{table}[h!]
    \centering
    \caption{The characterization of 6T SRAM, 2T eDRAM, and our MCAIMem. The experiment of 1MB memory designs.}
    \label{tab:compare_memory}
    \resizebox{\columnwidth}{!}{
    \begin{tabular}{|l|c|c|c|c|}
    \hline
         \multirow{1}{*}{eRAM types} & SRAM & eDRAM(2T) & \makecell{MCAIMem}\\ \hline
         \makecell{Static \\power (mW)} & 19.29 & \makecell{0.84 (Min)\\5.03 (Max)} & \makecell{3.15 (Min)\\ 6.82 (Max)}  \\ \hline
         Read (pJ) & 0.08 & \makecell{0.00016 (Min)\\0.14 (Max)} & \makecell{0.01014 (Min)\\ 0.1325 (Max)} \\ \hline
         Write (pJ) & 0.16 & \makecell{0.00016 (Min)\\ 0.0184 (Max)} & \makecell{0.02014 (Min)\\0.0361 (Max)}  \\ \hline
    \end{tabular}}
\end{table}

As depicted in Fig.~\ref{fig:layout}, the mixed SRAM and eDRAM design reduces the area size by 48\% compared to SRAM memory alone. During circuit simulation, the asymmetric characteristic of 2T eDRAM causes the stored data value to impact static and access power. When all bit data is 1, eDRAM consumes less power as the leakage is substantially lower when the storage node is at VDD. The gate leakage from VDD to the storage node is minimal, with the main leakage now being the subthreshold current of the PMOS. This current is small, as we apply a delta of 0.4V to the gate of the PMOS access transistor. If all bit data is 0, a higher gate leakage current from VDD attempts to recharge the storage nodes to bit-1. Consequently, the one-enhancement technique increases the number of bit-1 bits, which is essential for reducing static power in 2T eDRAM.

Compared to 2T eDRAM, our mixed-cell memory includes one 6T SRAM and seven 2T eDRAMs. Static power originates from both the SRAM and eDRAM, yet it can be reduced by 3-6$\times$ compared to SRAM alone. Regarding read and write operations, the 6T SRAM is mostly balanced, while 2T eDRAM continues to display asymmetric characteristics. When reading a bit-1, the initial BL is VDD, so there is no change in the sense amplifier, leading to low energy consumption. Conversely, when reading a bit-0, the storage node must be recharged to 0, with the current from the storage node being the main contributor to energy consumption. 

\subsection{System evaluation}

This evaluation seeks to determine MCAImem's influence on DNN applications by simulating diverse CNN networks such as LeNet, VGG11, VGG16, AlexNet, and ResNet-50 with datasets including MNIST, CIFAR10/100, and ImageNet. In addition, we run simulations of I-BERT for language networks and CycleGAN for generative networks. We have modified SCALE-Sim~\cite{samajdar2018scale} to estimate the static and dynamic energy consumption of each memory device, considering the configurations of Eyeriss~\cite{chen2016eyeriss} and Google TPUv1~\cite{chen2016eyeriss}. In order to adapt our power model to various device configurations, adjustments were made according to their memory requirements. Specifically, for Eyeriss, which demands 108KB of SRAM, we modified the embedded RAM power model by reducing it to one-tenth of our original 1MB memory device configuration. Conversely, for Google TPUv1, which necessitates 8MB, we augmented the embedded RAM power model by a factor of eight. Regarding the RRAM model, we employed the model found in ~\cite{prabhu2022chimera}, operating under the assumption that both weight and activation utilize the RRAM as the on-chip buffer. This reflects memory size adaptations pertinent to both Eyeriss and TPUv1 configurations. Furthermore, we attribute no static power to RRAM, given that its non-volatile memory can toggle on and off without data loss, focusing our considerations solely on the read and write energy per byte.

\begin{figure}[!t]
\centering
\includegraphics[width=\linewidth]{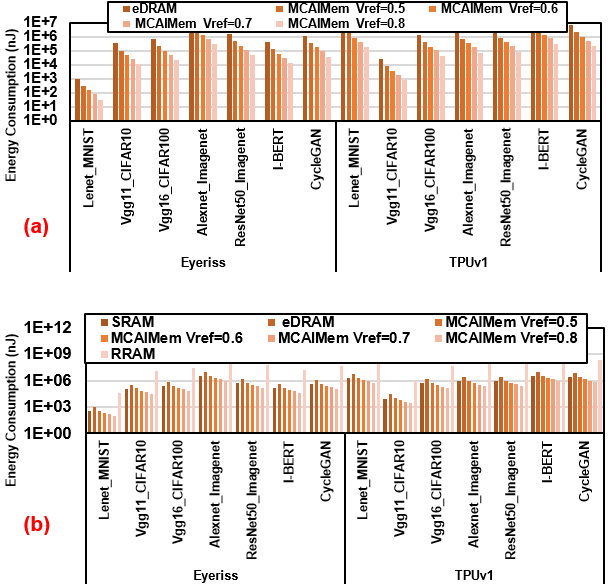}
\caption{(a) Comparison of required refresh energy consumption: 2T eDRAM vs. MCAIMem with different $V_{REF}$. (b) Comparison of total energy consumption: SRAM, RRAM, eDRAM, and MCAIMem with various DNN benchmarks on Eyeriss and Google TPUv1}
\label{fig:totalenergy}
\end{figure}

In this simulation-centric study, we extract the computation time for each device configuration, operating under an assumed clock frequency of 100 MHz. Subsequent to determining the computation time for each memory type, we apply respective power models to calculate the final static and dynamic energy. The 6T SRAM and conventional 2T eDRAM (sans the one-enhancement encoder/decoder) are utilized as our baseline comparison. Our evaluation is meticulously confined to the on-chip buffer performance, intentionally omitting the energy associated with MAC operations. We've opted for a clock frequency of 100MHz, in alignment with the slowest operational clock frequencies observed in AI accelerators—Eyeriss at 100MHz~\cite{chen2016eyeriss} and TPUv1 at 700MHz~\cite{jouppi2021ten} serve as pertinent examples. This chosen clock frequency not only determines the time to access and hold memory across layers but also proves imperative given our usage of eDRAM, which necessitates a refresh operation to safeguard data. The clock frequency thereby becomes vital in estimating the requisite number of refresh operations amidst AI accelerator computations. Employing SCALE-SIM enables us to quantify the number of clock cycles, and owing to the systolic array's design, each clock cycle concurrently facilitates MAC and memory access, thereby simplifying the tally of on-chip memory accesses. Fig.~\ref{fig:totalenergy} delineates the minimum power savings attainable. Interestingly, with the adoption of a swifter clock frequency, data retention time per layer is truncated, possibly culminating in diminished power savings due to a decrease in the number of refresh operations while computations are in progress.

\begin{figure}[!t]
\centering
\includegraphics[width=\linewidth]{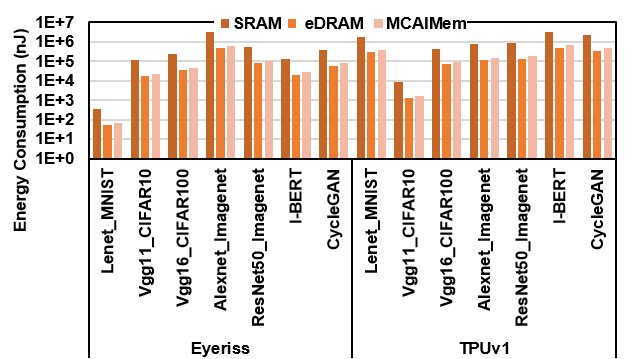}
\caption{The static energy consumption comparison: analyzing 6T SRAM, 2T eDRAM, and MCAIMem across various deep learning networks. Evaluation is on Eyeriss and TPUv1 platforms for a comprehensive assessment.}
\label{fig:staticpower}
\end{figure}

When it comes to static power, SRAM energy consumption is higher than both the 2T eDRAM and our mixed-cell memory. Although our mixed-cell memory has a higher static energy consumption than the 2T eDRAM, it performs better than SRAM. With a 1/7 SRAM/eDRAM ratio, the fixed energy overhead of SRAM in our mixed-cell memory accounts for 76.5\% of the total consumption. Further details can be found in Fig.~\ref{fig:staticpower}.

As for refresh power, SRAM does not require refresh operations, while 2T eDRAM and our mixed-cell memory do. Adjusting the reference voltage ($V_{REF}$), as discussed in Section~\ref{sec:refreshextention}, can help increase the refresh period and reduce refresh operations. We choose conventional 2T eDRAM with current-mode sense amplifier and experiment with $V_{REF}$ values of [0.5, 0.6, 0.7, 0.8] for voltage-mode sense amplifier in our mixed-cell memory. Fig.\ref{fig:totalenergy}.(a) demonstrates that refresh energy can be significantly reduced with the proper $V_{REF}$ value. Consequently, our chosen $V_{REF}$ of 0.8 yields the lowest refresh operation since it extends the refresh period nearly 10$\times$, from 1.3us to 12.57us.

Regarding total energy consumption, which encompasses both static and dynamic energy utilization throughout the inference process, eDRAM delivers a compact area footprint yet does not excel in overarching energy consumption due to its refresh energy requirements. Contrarily, our mixed-cell memory affords advantages in both area footprint minimization and energy consumption reduction, attaining an energy efficiency that is 3.4$\times$ superior to 6T SRAM, as demonstrated in Fig.~\ref{fig:totalenergy}.(b). Nevertheless, RRAM lags in energy efficiency, being over 100X higher than SRAM, attributed to its requisite for substantial write operations.


\begin{figure}[!t]
\centering
\includegraphics[width=\linewidth]{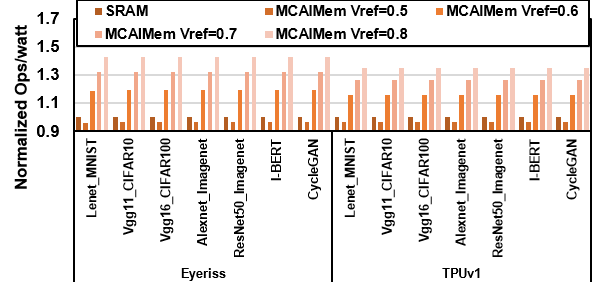}
\vspace{-2em}
\caption{Normalization of Ops per watt improvement based on the SRAM and MCAIMem configurations across different DNN benchmarks on Eyeriss and Google TPUv1.}
\label{fig:normalOpsPerWatt}
\vspace{-2em}
\end{figure}

Given that the on-chip buffer contributes to 42.5\% of the power consumption in Eyeriss~\cite{chen2016eyeriss} and 37\% in TPUv1~\cite{jouppi2017datacenter}, utilizing an MCAIMem configuration with $V_{REF}$=0.8 allows our performance-per-watt to attain gains between 35.4\% and a peak of 43.2\%, surpassing the efficiency of an on-chip buffer that employs SRAM, as depicted in Fig.~\ref{fig:normalOpsPerWatt}. Consequently, MCAIMem emerges as a compelling solution, potentially paving the way for innovations in efficient AI memory design.

\section{Related works}
\label{sec:relatedworks}

Deep Neural Networks (DNNs) require a vast number of parameters to achieve superior performance, leading to increased memory demands. Addressing the increased requirements for the on-chip data buffer and data movement is essential for enhancing DNN accelerator performance. Chen et al.~\cite{chen2016eyeriss} demonstrated that off-chip DRAM access consumes 200$\times$ more energy and has a longer access time compared to ALU. As a result, optimizing the on-chip buffer has become a primary challenge in boosting DNN accelerator throughput. The key question is how to maximize on-chip memory capacity and minimize off-chip access to improve energy efficiency in DNN accelerators.

DaDianNao~\cite{chen2014dadiannao} suggests replacing SRAM with fully eDRAM (1T1C) in conventional DNN accelerators to increase on-chip buffer capacity significantly. However, this approach necessitates periodic refreshes to maintain DNN data, leading to substantial energy consumption—accounting for 38.3\% of the total DNN accelerator power usage. RANA~\cite{tu2018rana}, a more recent technique, exploits the shorter activation data lifetime compared to eDRAM retention time, allowing the elimination of unnecessary refresh operations. As DNN applications evolve, this observation may become less applicable, resulting in increased activation data and potential violations of the activation data lifetime constraint.

Computing-in-memory (CIM) has been proposed as an alternative to traditional DNN accelerators, aiming for improved throughput. Techniques such as the 4T Dual~\cite{yoo2019logic} eDRAM array, which employs two 2T eDRAMs, and DualPIM~\cite{jung2022dualpim}, which utilizes hybrid SRAM and eDRAM configurations as computation nodes, have been developed. Additionally, recent eDRAM~\cite{ha202236} node optimization focuses on reducing leakage and enhancing robustness in CIM. While these approaches demonstrate substantial performance and energy savings, the need for on-chip buffers remains.

Furthermore, a study called ZEM~\cite{nguyen2021zem} explores the asymmetric characteristics of DNN data to extend DNN data retention time in off-chip DRAM, significantly reducing off-chip DRAM power. However, this work primarily aims to decrease off-chip DRAM power during DNN application processing rather than tackling the core challenge of improving DNN accelerator performance by minimizing off-chip DRAM access. In this paper, we propose a design that targets the on-chip buffer by creating a mixed SRAM and eDRAM cell design to minimize the area and energy consumption of the on-chip buffer. This approach holds promise for the design of on-chip buffers in next-generation DNN accelerators.

\section{Conclusion}
\label{sec:conlusion}

In this paper, we introduce MCAIMem, an innovative area and energy-efficient AI memory design that utilizes a mixed CMOS memory cell design, comprising both SRAM and eDRAM cells. We optimize the ratio of SRAM/eDRAM cells to achieve reduced area and capitalize on DNN's data representation and asymmetric eDRAM cells for lower energy consumption. Experimental results demonstrate that our MCAIMem design can decrease the area by 48\% and energy consumption by 3.4$\times$ compared to conventional SRAM designs, without sacrificing accuracy. This work highlights the potential of mixed CMOS memory cells and asymmetric 2T eDRAM cell implementations in attaining an optimized balance between performance, area, and energy consumption for AI memory designs. In conclusion, our mixed CMOS cell memory design, MCAIMem, provides a promising solution and has the potential to become a new standard for efficient AI memory design.

\section*{Acknowledgment}
This work was supported in part by CoCoSys, a JUMP2.0 center sponsored by DARPA and SRC, the National Science Foundation (CAREER Award, Grant \#2312366, Grant \#2318152), TII (Abu Dhabi), and the DoE MMICC center SEA-CROGS (Award \#DE-SC0023198).

\bibliographystyle{IEEEtranS}
\bibliography{main}

\end{document}